**Detection of an excess of young stars in the Galactic center Sagittarius B1 region.**

F. Nogueras-Lara[1], R. Schödel[2], and N. Neumayer[1]

[1]Max-Planck Institute for Astronomy, Königstuhl 17, 69117 Heidelberg, Germany
[2]Instituto de Astrofísica de Andalucía (CSIC), Glorieta de la Astronomía s/n, 18008 Granada, Spain
email: nogueras@mpia.de

The Milky Way's center is the closest galaxy nucleus and the most extreme environment of the Galaxy. Although its volume is less than 1% of that of the Galactic disk, up to 10% of all new-born stars in the Galaxy in the past 100 Myr formed there. Therefore, it constitutes a perfect laboratory to understand star formation under extreme conditions, similar to those in starburst or high-redshift galaxies. However, the only two known Galactic center young clusters account for <10% of the expected young stellar mass. We analyze the star formation history of Sagittarius (Sgr) B1, a Galactic center region associated with strong HII emission, and find evidence for the presence of several $10^5 M_\odot$ of young stars, that formed ~10 Myr ago. We also detect the presence of intermediate age stars (2-7 Gyr) in Sgr B1 that appear to be rare (or absent) in the inner regions of the nuclear stellar disk, and might indicate inside out formation. Our results constitute a large step toward a better understanding of star formation at the Galactic center, such as the fate of young clusters, and the possibly different initial mass function in this region.

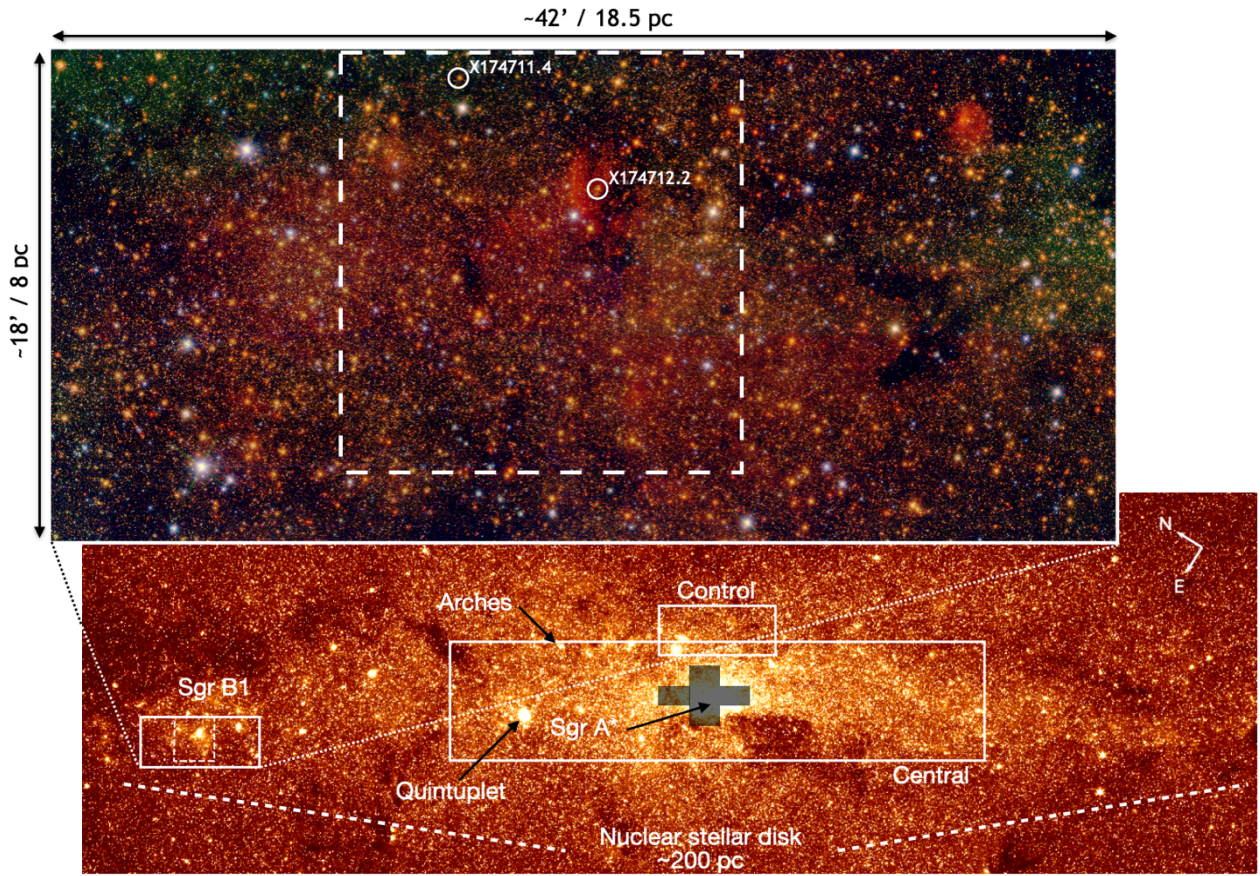

Fig. 1: *Scheme of the analyzed regions (Sgr B1, control, and central nuclear disk) plotted over a Spitzer 4.5 μm image[34]. The black shaded cross indicates a region of low completeness dominated by the nuclear star cluster that was excluded from the analysis of the central field[4]. The position of the Arches and Quintuplet clusters, and Sgr A\* are indicated. The Zoom-in region corresponds to a JHKs GALACTICNUCLEUS false color image. The dashed rectangle indicates the region of intense hot dust emission that was specifically studied. Two Wolf-Rayet stars[11] within the analyzed region are indicated.*

The Milky Way's center is located at only 8 kpc from Earth, and constitutes the only galaxy nucleus where we can resolve individual stars down to milli-parsec scales[1,2]. It is roughly delimited by the central molecular zone and the nuclear stellar disk, a dense stellar disk-like structure[3,4]. The nuclear stellar disk is characterized by high stellar densities, strong tidal fields, high magnetic fields, and high turbulence and temperature of the interstellar medium[3,5,6]. In spite of — or possibly because of — these harsh conditions, the Galactic center emits more than 10 % of the total Galactic Lyman continuum flux in spite of occupying less than 1 % of the total Galaxy volume[7]. Studies of the radio to high energy emission, the finding of massive young stars throughout its region, the detection of classical Cepheids, as well as the star formation history inferred from luminosity functions, all indicate that the star formation reached values of the order of 0.1 $M_\odot$/yr, in the Galactic center in the past 10-100 Myr[4,8,9,10,11]. Therefore, it constitutes a unique laboratory to study star formation under extreme conditions. However, there are only two young massive clusters known (Arches and Quintuplet), that account for less than 10% of the total young stellar mass expected[12]. This is the so-called missing clusters problem. A plausible explanation for this is the rapid dissolution of even the most massive clusters, due to the tidal field in the Galactic center and encounters of the young clusters with massive molecular clouds[13,14]. Moreover, the high stellar background density[2] combined with the strong and patchy interstellar extinction toward the Galactic center[15] hamper the detection of all but the tightest and most massive young clusters as well as of individual young stars, and restrict their analysis to mainly the near infrared regime. Therefore, the creation of a complete census of recent star formation in this region is a formidable challenge.

Sgr B1 is a well-known region associated with strong HII emission in the nuclear stellar disk[16]. Far-infrared observations suggest the presence of widely spaced hot stars that excite the gas in at least eight separate sub-regions[16,17]. Moreover, a cohort of 6 young massive stars (O-type and WN7-9ha) have been detected there[11,18]. In this work, we use the GALACTICNUCLEUS survey[1,2] —a high-angular resolution (~0.2'') JHKs catalog specifically designed to observe the Galactic center— to study a field of ~160 pc$^2$ covering part of the Sgr B1 region (centered on 17$h$ 47$m$ 15.41$s$ -28° 31' 39.7', Fig. 1), and compare it with a control field of similar size in the inner nuclear stellar disk region (centered on 17$h$ 45$m$ 20.81$s$ -28° 57' 58.6', Fig. 1). We chose this control field because it was observed under similar excellent conditions as the target field[2] (seeing in H, Ks ~0.4''), and because it does not contain any obvious structures, such as the Arches or Quintuplet clusters or the nuclear star cluster.

Given the significantly different extinction along the line of sight between the Galactic center and the Galactic disk, we applied a color cut in the HKs color-magnitude-diagram to remove foreground stars[4,19] (Fig. 2a). We then built extinction maps using red giant stars —whose intrinsic color $(H-Ks)_0$ is approximately constant— to correct for reddening and differential extinction[15]. Finally, we created a Ks luminosity function (KLF, Fig. 2b, corrected for completeness), that gives the number of stars per luminosity interval.

The KLF of a stellar population contains information about its formation history[4,20]. We fitted the KLFs of the Sgr B1 and the control fields with a linear combination of theoretical models applying Monte Carlo (MC) simulations (see Methods). The KLF of a single-age population changes as a function of age, with the variability time-scales shortening towards younger ages. We chose the age-sampling particularly closely spaced for the youngest ages (5-40 Myr). We used two different sets of stellar evolutionary models that properly cover the youngest stellar populations, to deal with possible systematics (Parsec[21,22,23] and MIST[24,25,26]). We assumed a metallicity of twice solar in agreement with recent results for the Galactic center[4,27,28].

**Results**

Figure 2 (c, d) shows the obtained star formation history (SFH) for the Sgr B1 and the control field. We also applied the same technique to a KLF of the central region of the nuclear stellar disk directly taken from a previous study[4] (~1600 pc$^2$, indicated in Fig. 1) to compare the SFHs (Fig. 2e). For this comparison we excluded the nuclear star cluster, as it exhibits a significantly different SFH[4,20,29] and might thus bias the results. The results for the control and the inner Galactic center fields agree with previous work for the central region of the nuclear stellar disk[4]: The bulk of stars (≳ 80 % of the stellar mass) in both fields is older than 7 Gyr. There was a distinct epoch of star formation between 0.5-2 Gyr ago, where we found a significant contribution from the stellar models with ages of ~1 Gyr, corresponding to the massive star forming event that took place ~1 Gyr ago in the nuclear stellar disk[4].

Star formation continued at lower level until the present moment. On the other hand, the SFH of the Sgr B1 region is significantly different. The stellar population is younger on average and there is an important contribution from an intermediate age (2-7 Gyr) population (~40 % of the total stellar mass), suggesting a more continuous SFH than in the innermost regions of the Galactic center. Moreover, the star formation activity between 0.5-2 Gyr is more prominent. Finally, we detect a much larger contribution from young stars in comparison to the control field. In particular, the youngest age bin (<60 Myr) accounts for more than 5 % of the total stellar mass of the Sgr B1 region, six times more than in the control field. This contribution from young stars is also around twice higher than in the central region of the nuclear disk, where we detect a higher fraction of young stars than in the control field due to the presence of the Arches and Quintuplet clusters, as well as probably unknown stellar associations in this large field[30].

On the other hand, the presence of a significantly different contribution from the intermediate age stellar population in the Sgr B1 region, in comparison with the control and the inner Galactic center fields, may be indicative of an inside-out formation of the nuclear stellar disk. Such an age gradient has also been observed in external galaxy nuclei and used to propose the inside-out formation channel of nuclear stellar disks[31].

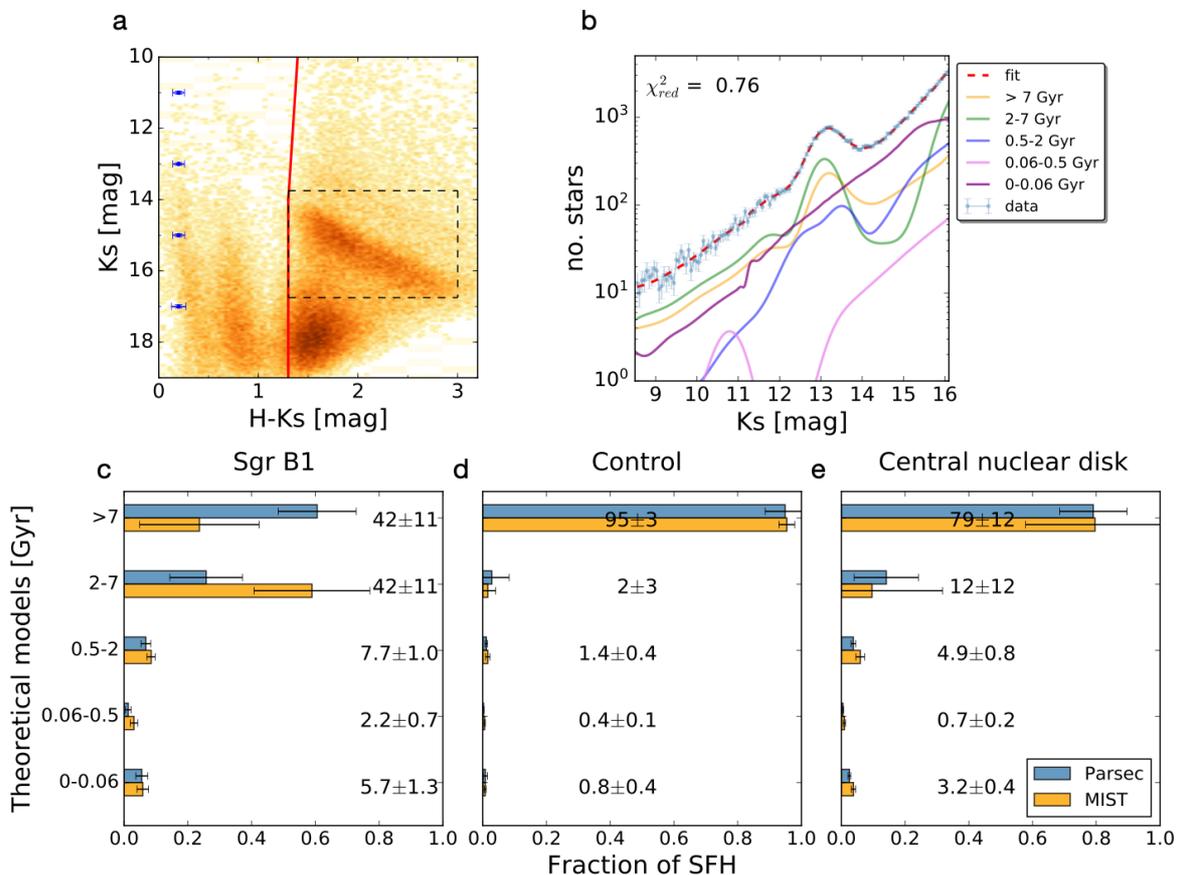

Fig. 2: **a**, H-Ks color-magnitude diagram (Vega magnitudes) for the Sgr B1 field. The red line shows the color cut to remove foreground stars. The black dashed rectangle indicates the red giant stars used to compute the extinction map. The blue error bars show the $1\sigma$ mean uncertainties of the data. **b**, Sgr B1 KLF (in blue) corrected for extinction, completeness, and saturation. The error bars show the $1\sigma$ uncertainties per brightness bin. The red dashed line indicates the best-fit model (Parsec models), whose reduced $\chi^2$ is indicated in the Figure. Colored lines depict the contribution of the different age bins. **c, d, e,** SFHs of the Sgr B1, the control, and the central region of the nuclear stellar disk. The error bars depict the $1\sigma$ uncertainties. The numbers in each panel indicate the percentage of total stellar mass per age interval computed as an average between Parsec and MIST models (see Methods V).

Using Parsec models, we estimate that the Sgr B1 field studied here contains a total originally created stellar mass of $(7.6\pm0.7) \cdot 10^6 \, M_\odot$. This implies that the youngest stellar population accounts for $(4.3\pm1.1) \cdot 10^5 \, M_\odot$, nearly an order of magnitude higher than the combined mass of the Arches and Quintuplet clusters[32,33].

We further investigated the presence of young stars in Sgr B1 by carrying out a dedicated analysis of a small region (~40 pc$^2$, white dashed rectangle in Fig. 1), that contains intense dust emission in the Spitzer image[34] (4.5 $\mu m$), probably caused by young hot stars. We found that more than 7 % of the total stellar mass is due to young stars (<60 Myr). We also estimated the age of the stars in this region by analyzing the contribution of the considered young stellar models (5, 10, 20, and 40 Myr) to each of the MC samples (see Methods). Using Parsec and MIST models, we found that the 5 and 10 Myr model populations significantly contribute to ~99% of all MC samples, and correspond to 6±1 % of the total stellar mass (Fig. 3), accounting for almost the full mass of young stars (0-60 Myr) in the region. Therefore, we estimate that this region contains ~$1.2 \cdot 10^5 \, M_\odot$ of stars with ages between 5-10 Myr. A similar analysis carried out for the whole Sgr B1 field indicates that 2±2 % of the total stellar mass is due to young stars with ages between 5-10 Myr. We thus conclude that the ~40 pc$^2$ region with intense hot dust emission presents an overabundance of young stars of these ages in comparison with the surrounding area.

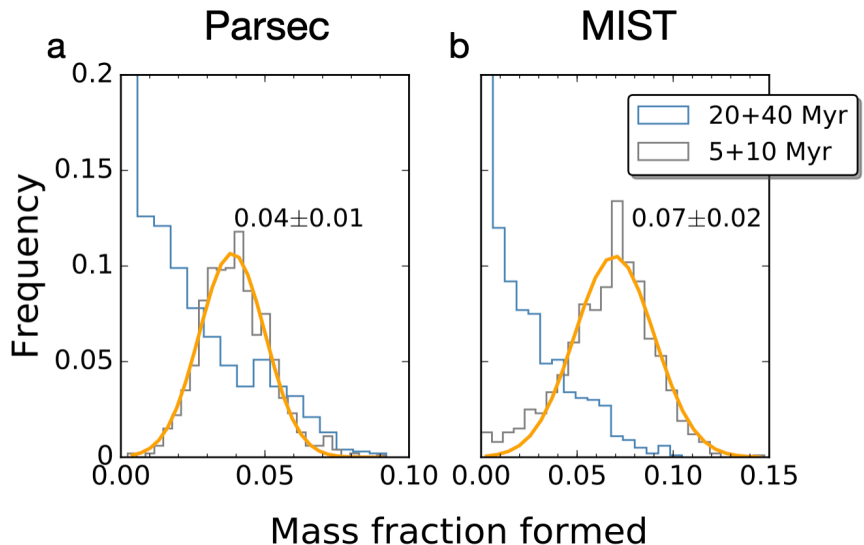

*Fig. 3: Young populations' contribution to the KLF of the Sgr B1 hot dust emission region using Parsec (a) and MIST (b) models. The yellow lines show a Gaussian fit to the distribution of 5+10 Myr. The mean and the standard deviation (1σ uncertainty) are indicated in each panel.*

The Sgr B1 region is located at a radial distance of ~80 pc from the supermassive black hole Sgr A*. This implies a rotation period of ~5 Myr for its stellar population, assuming a circular velocity of ~100 km/s ref [35]. Thus, our age estimate for the detected young stellar population suggests that it was not formed in-situ and has already orbited the nuclear stellar disk at least once. This agrees with previous work proposing that the exciting sources of Sgr B1 did not originate there[16,17,36]. Hence, the detected HII emission is a consequence of the ionization of the envelope of molecular gas and dust found in the Sgr B complex by stars widely spread throughout the field[16]. This scenario also explains the detection of some young stellar objects in the Sgr B1 region as a very recent star formation event (different from the detected young stellar population), triggered by outflows from the ionizing stars[16].

On the other hand, the presence of a cohort of 6 O-type and WN7-9ha X-rays emitter stars in this region with apparent co-eval evolution and relative proximity between them, suggests that they constitute a distinct physically related group[18]. A comparison of such stellar population with the Arches cluster, where only 4 members are known X-rays emitters[37], suggests that the Sgr B1 cohort of known hot massive stars could be representative of a similarly rich massive stellar population[18], in good agreement with our findings. In addition, the presence of a supernova remnant candidate in the Sgr B1 region[38], near the detected young stellar association (~15 pc away from it), gives further evidence for the presence of a young stellar population, in agreement with our findings.

Determining the location where the detected young stellar population was formed is very challenging due to its age uncertainty and the unknown distance along the line-of-sight that impedes an accurate orbit reconstruction. Given the estimated high mass of the young stellar population (~$10^5 M_\odot$), it seems unfeasible that the young stars were formed as a single bound cluster. Moreover, theoretical studies point toward an upper limit of ~$10^4 M_\odot$ for the formation of bound clusters in the central molecular zone[39]. Therefore, we conclude that the detected young stellar population probably formed as a co-eval stellar association. Nevertheless, extrapolating the cluster formation efficiency observed in the close Galactic center cloud Sgr B2 (where ~40 % of the stars are forming in gravitationally bound clusters[40]), it is likely that some of the young stars in Sgr B1 could have constituted several gravitationally bound clusters. This can be used to estimate a lower limit on the age of the detected stellar population, given that there is not any clear stellar over-density in the Sgr B1 region, pointing towards a rapid dispersion of the possible young clusters below the high stellar background density in the Galactic center. In this way, the age that we estimate for the detected young stellar association is consistent with the predicted short disruption time of gravitationally bound clusters (~6 Myr), caused by tidal shocking by giant molecular clouds in the central molecular zone[14]. Moreover, it also agrees with the previously inferred recent star formation history in the Galactic center that presents a maximum of star formation around 10 Myr ago[41].

To our knowledge, this is the first detection of a significant mass of recently formed stars at the Galactic center beyond the Arches and Quintuplet clusters and the known isolated massive stars. Our findings may indicate the fate of the Arches and Quintuplet clusters, that are around 5 million years younger than the detected young stellar population. In this way, they contribute to a more general picture of the evolution of the young stars in the Galactic center in which stars form in massive stellar associations that can contain clusters (Sgr B2 is an example of this stage[40]), and later get dispersed while orbiting through the nuclear stellar disk. Our results also help understand the isolated massive stars detected across the Galactic center, whose proper motions indicate they are not related to the so-far known young clusters[42], supporting their formation in stellar associations or gravitationally bound clusters that are dispersed on rather short timescales after their formation several million years ago.

**Methods**

**Data Availability**

All the raw data used in this study are available at the ESO Science Archive Facility (http://archive.eso.org/eso/eso_archive_main.html) under program ID 091.B-0418. The final version of the GALACTICNUCLEUS survey (images and point source catalogues) is publicly available via the ESO phase 3 data release interface (https://www.eso.org/qi/catalogQuery/index/369).

**Code Availability**

The described analysis of the luminosity function was performed using the Python *Scipy* package (https://scipy.org). The used Parsec[21,22,23] and MIST[24,25,26] models are publicly available online (http://stev.oapd.inaf.it/cgi-bin/cmd and https://waps.cfa.harvard.edu/MIST/, respectively). The main codes used for the analysis are publicly available at https://github.com/fnogueras/Codes_GC_young_stars_excess. Other scripts corresponding to particular steps of the analysis, can be obtained from the corresponding author upon reasonable request.

## I Data

For this work we used H and Ks data from the GALACTICNUCLEUS survey[1,2], that is publicly available on the ESO Phase 3 data release archive. This is a high-angular resolution (~0.2'') JHKs survey of the Galactic center especially designed to observe its stellar population overcoming the extreme extinction and source crowding. It contains accurate photometry of ~$3.3 \cdot 10^6$ stars covering a total area of ~6000 pc$^2$. The photometric statistical uncertainties are below 0.05 mag at H~19 and Ks~18 mag. The zero-point systematic uncertainty is ≲ 0.04 mag in all bands. In particular, we used two individual pointings (D12 and F19)[2], partially covering the Sgr B1 region and a control field, that were observed under similar excellent conditions (seeing in H, Ks ~0.4''). As comparing fields, we also used data from pointing F10, containing the Quintuplet cluster, and the 14 central pointings of the survey that cover the central region of the nuclear stellar disk[4]. For the latter case, we directly used the final KLF as it was obtained in previous work[4], where the nuclear star cluster — that constitutes a distinct component with different origin, SFH, and stellar population than the nuclear stellar disc— was removed to not bias the results. The observing conditions of these 14 fields were somewhat worse than the ones for the target and control field (seeing in H, Ks ~0.6''), and thus the faint end of the KLF is ~1 mag lower. Moreover, the central region of the GALACTICNUCLEUS survey has a low completeness for its innermost field (cross-shaped region in Fig. 1a), impeding a proper analysis of the KLF of the NSC to derive its SFH.

Prior to our analysis, we corrected potential saturation problems in Ks for stars brighter than 11.5 mag[2]. For this, we used the SIRIUS IRSF[43] survey of the Galactic center to replace the photometry of saturated stars, and also completed the list with bright stars that might have escaped detection in the GALACTICNUCLEUS catalogue.

## II Extinction maps

For each of the analyzed regions, we created a dedicated extinction map using red clump stars[44] (red giant stars in their helium burning phase), that are very abundant and homogeneously distributed across the field, and have a well-defined intrinsic color[15] $(H-Ks)_0=0.10\pm0.01$ mag. We also included red giant stars (with very similar intrinsic colors H-Ks)[4,15], to increase the angular resolution of the maps. To choose the reference stars, we used a color cut in the color-magnitude diagrams as shown in Fig. 2a (black dashed rectangle). We built the extinction maps following the methodology described in our previous work[15] and assuming an extinction curve[45] $A_H/A_{Ks} = 1.84\pm0.03$. We defined a pixel size of ~2'', and computed the associated extinction values for each pixel by using the five closest reference stars in a maximum radius of 7.5''. We weighted the distances using an inverse-distance weight method, and assumed a maximum color difference of 0.3 mag between the stars to avoid mixing stars with too different extinction. If less than five reference stars were detected for a given pixel, we did not assign any extinction value ("NaN"). Figure 4 shows the extinction maps obtained for the Sgr B1 and the control field (F19). Using a Jackknife resampling method, systematically leaving out one of the reference stars for each pixel, we obtained a statistical uncertainty of the extinction maps of ~3 %. The systematics were estimated quadratically propagating the uncertainties of the quantities involved in the extinction calculation. We obtained a mean systematic uncertainty of ~5 %.

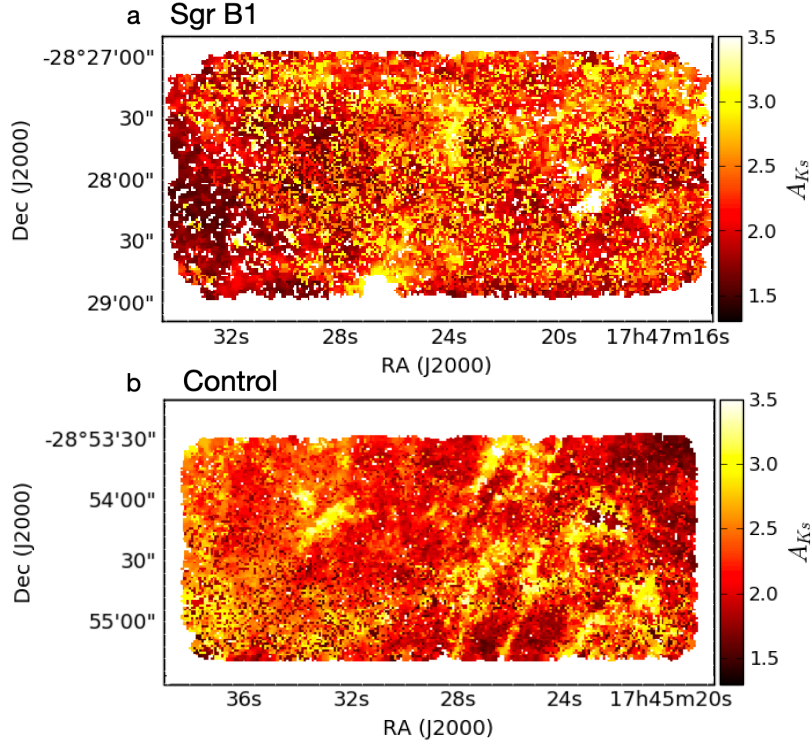

*Fig. 4: Extinction maps for Sgr B1 (a) and the control field (b). White pixels indicate "NaN" values, meaning that there were not enough reference stars to compute an associated extinction value.*

## III De-reddening

Given the extreme differential extinction along the observed line of sight, it is possible to remove the foreground stellar population —belonging to the Galactic disk and bulge— by applying a color cut H-Ks~1.3 mag[19] (red solid line in Fig. 2a). We then applied the previously computed extinction maps to deredden each of the studied fields. Figure 5 shows the color-magnitude diagrams H-Ks of the Sgr B1 field and the control region before and after applying the extinction maps. To check that the differential extinction is significantly corrected, we computed the standard deviation of the distribution of red clump stars before and after the extinction correction, and obtained that the scatter of red clump stars is ~8 times lower in the corrected sample.

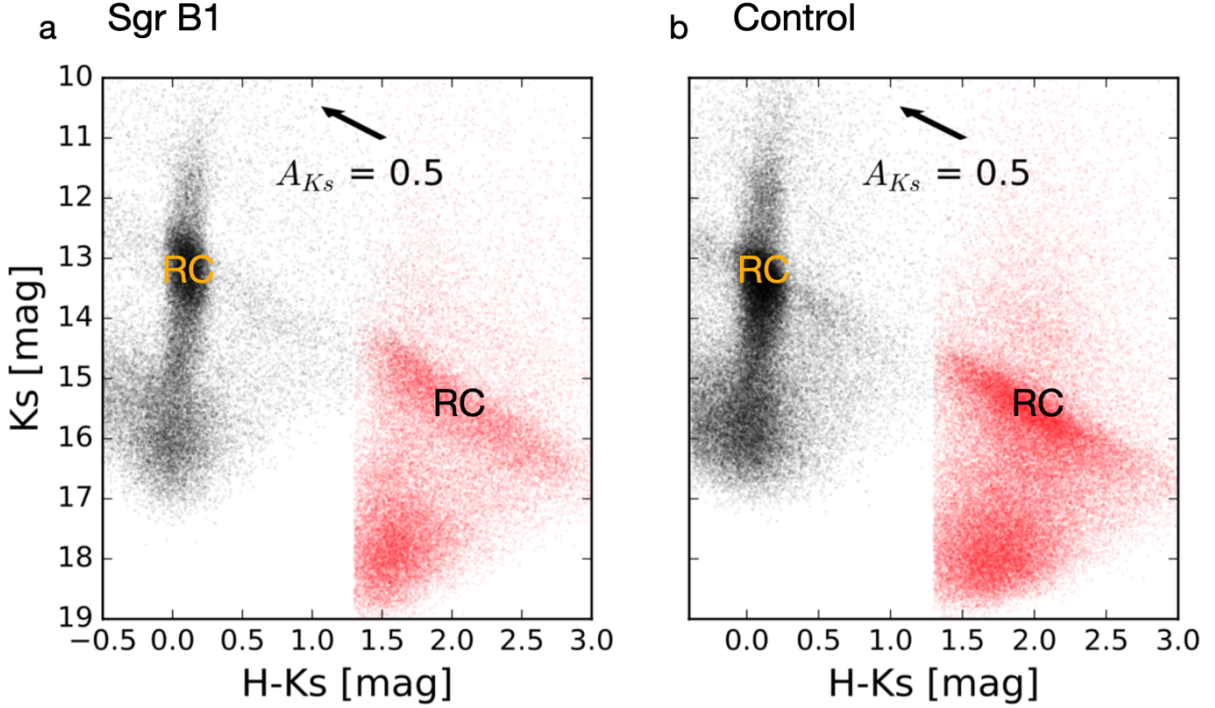

*Fig. 5: Color-magnitude diagrams Ks versus H-Ks (Vega magnitude system) before (red dots) and after applying the extinction maps (black dots) for the Sgr B1 (a) field and the control region (b). The black arrows in the panels indicate the reddening vector for an extinction in Ks of 0.5 mag. The position of the red clump (RC) is indicated before and after the correction.*

### IV Ks Luminosity Function

To create the KLFs, we used all the Ks de-reddened stars (including stars that were not detected in H-band)[4], after excluding the foreground population. We also removed over-de-reddened stars by excluding stars with a de-reddened H-Ks color $2\sigma$ bluer than the one from the mean distribution of the de-reddened red clump feature. Around 3% of the stars were removed by applying this technique for the analyzed Sgr B1 region. They correspond to stars with an average H-Ks ~ 1.6 mag that is considerably lower than the mean value of the whole stellar population (H-Ks ~ 2 mag). We conclude that they are likely foreground stars from the Galactic bulge that passed the previous color cut, but are not still behind the full extinction screen for the Galactic center.

We created the KLF assuming the bin width that maximizes the Freedman-Diaconis[46] and Sturges[47] estimators. We computed the uncertainties considering Poisson errors (i.e. the square root of the number of stars per bin).

*Completeness*

We also corrected for completeness computing a solution based on artificial stars tests. Namely, we created 20 modified science images for each field inserting ~5 % of the total number of stars, in magnitude bins of 0.5 mag starting from Ks=12 mag. We then used the StarFinder software[48] to measure point spread function photometry following the same procedure used to create the GALACTICNUCLEUS survey[1,2], and checked the fraction of recovered artificial stars to estimate the completeness correction. Figure 6 indicates the completeness solution for the Sgr B1 and the control fields. The uncertainties were estimated via the standard deviation of the completeness solution of the results obtained for each of the four independent HAWK-I chips that constitute each field[1]. We found that the completeness is higher for the Sgr B1 region, which can be explained due to the higher stellar crowding in the control field that is closer to the innermost regions of the Galactic center.

We computed an extinction correction for each completeness solution by calculating the median extinction of the stars constituting each KLF. We then completed the KLFs setting a lower limit of 75% of data completeness. We estimated the uncertainty per magnitude bin quadratically propagating the uncertainties from the completeness solution and the original KLF[4].

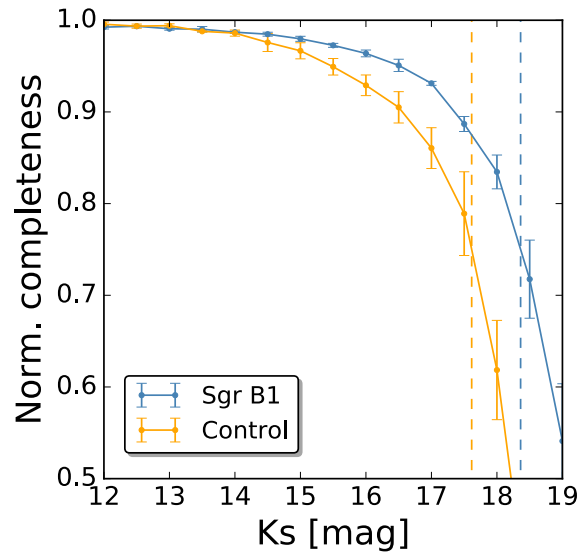

*Fig. 6: Completeness solution for the Sgr B1 region and the control field. The 1σ uncertainty is indicated in each case. The vertical dashed lines show the 75 % level of completeness for each region. The KLFs are not corrected for extinction.*

*Saturation*

We restricted the analysis of the dereddened KLF to Ks>8.5 mag to avoid problems due to potential saturation of bright stars in the SIRIUS IRSF catalogues[49]. We tested this saturation limit using 2MASS data[50], whose lower angular resolution (~2'') makes it ideal to observe bright stars without saturation. We used 2MASS H and Ks photometry in the Sgr B1 region defined in this work and searched for common stars to stablish a common photometric zero point. We then removed foreground stars in the 2MASS data by using a color cut around H-Ks~1.3 mag, as previously explained. We applied the extinction map derived for the Sgr B1 region and created a Ks luminosity function corrected for extinction. Figure 7 shows the comparison between the 2MASS and the used KLFs. We conclude that, given the saturated sources in our sample that are potentially not detected, the KLF is significantly incomplete for Ks<8.5 mag.

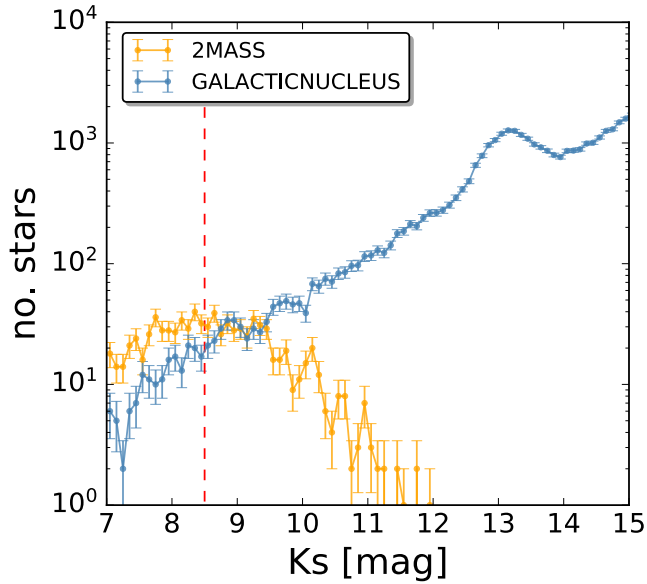

*Fig. 7: KLFs comparison between 2MASS and the used data for the Sgr B1 region. The error bars show the 1σ uncertainties per brightness bin. Both KLFs are corrected for extinction. The red dashed line shows the brightness cut (Ks = 8.5 mag) used to avoid incomplete brightness bins of the KLF due to saturation.*

**V Model fitting and SFH calculation**

The KLF contains fundamental information about the SFH that can be reconstructed studying its shape and the properties of its main features[4,20,29]. In this way, stellar populations with different ages present characteristic KLFs as it is shown in detail in Methods VI. The main features[51] observed in the KLF are the asymptotic giant branch bump (due to stars that at the beginning of helium shell burning asymptotic giant branch evolution), the red clump bump (caused by red giant stars burning He in their cores), the red giant branch bump (due to old stars whose H-burning shell approaches the composition discontinuity left by the deepest penetration of the convective envelope during the first dredge-up[52]), or the ascending giant branch (that contains stars evolving after the main sequence). Moreover, the presence of young stars (≲10 Myr) can be identified due to an increased number of counts fainter than the red clump, in contrast to older populations (see Methods VI), that appear as a consequence of bright main-sequence stars. In this way, we checked that the KLFs from the Sgr B1 and the control regions are significantly different (Fig. 8), pointing toward the presence of different stellar populations (and probably young stars, ≲10 Myr) as we later confirmed with a careful analysis of the KLF.

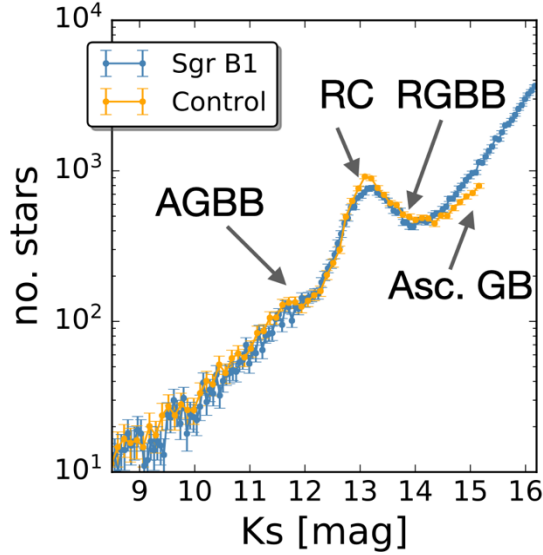

*Fig. 8: KLFs comparison between the Sgr B1 region and the control field. The KLFs are corrected for extinction. The control region KLF was scaled to the Sgr B1's one, for a visual comparison, given their different stellar masses and bin widths. The error bars correspond to 1σ uncertainties. The position of asymptotic giant branch bump (AGBB), the red clump feature (RC), the red giant branch bump (RGBB), and the ascending giant branch (Asc. GB) are shown in the figure. The RC and the Asc. GB features present significant deviation between both KLFs. The KLFs difference for magnitudes fainter than the RC points toward the presence of a young stellar population in the Sgr B1 region.*

We derived the SFH of each of the studied regions by fitting the KLFs with a linear combination of theoretical models[4,20]. To search for young stars that are potential tracers of dissolved clusters, we used PARSEC models[21,22,23] as a reference, given that they properly cover young stellar ages. We chose 14 individual ages for the model fit, that homogeneously sample the possible ages of the analyzed stellar populations: 14, 11, 8, 6, 3, 1.5, 0.6, 0.4, 0.2, 0.1, 0.04, 0.02, 0.01, 0.005 Gyr. We assumed a Kroupa initial mass function (IMF) corrected for unresolved binaries[53], and a stellar metallicity of around twice solar ($Z = 0.03$), in agreement with previous results for the Galactic center[4,27,28]. To fit the KLFs, we included a parameter to account for the distance modulus (~14.6 at the Galactic center distance), and allowed it to varied within $3\sigma$ of the quadratically propagated uncertainties of the distance and the uncertainty of the de-reddening process. We also included a Gaussian smoothing parameter to account for possible distance and/or differential extinction variations of the considered stellar populations.

To compute the SFH of each region, we resorted to Monte Carlo (MC) simulations creating 1000 KLFs obtained by randomly varying the number of stars per bin assuming the $1\sigma$ uncertainty as the standard deviation of the distribution. We then fitted each of the MC samples using a chi2 minimization criterion and obtained the SFH as the average of the results. To minimize possible degeneracies between models with similar age, we combined them into 5 final age-bins, as shown in Fig. 2c, d, e. The $1\sigma$ uncertainty is obtained via the standard deviation of the results in each age bin. To address potential systematic effects due to model selection, we repeated the procedure assuming MIST models[24,25,26], that are independently created and also adequate to trace the young stellar ages. We chose similar ages, a metallicity of around twice solar, and a Salpeter IMF. The final value for each of the 5 age bins was computed averaging over the results obtained with Parsec and MIST. We estimated the final uncertainties quadratically propagating the ones independently obtained by using Parsec and MIST models.

We fitted the KLFs computed for the Sgr B1 and the control regions, and also the KLF derived for the central region of the nuclear stellar disk in our previous work[4] (where the final uncertainties were larger due to the lower completeness of the KLF due to the worse data quality). We found a significantly different SFH in the Sgr B1 region, where there is a decrease of the contribution of old stars compensated by an increase of intermediate age stars, and a significant contribution of young stars, suggesting the presence of an association of young stars.

We analyzed potential sources of systematic uncertainties to assess the obtained results:

**(1) Stellar metallicity**. We used Parsec models assuming solar and 1.5 solar metallicity to assess the results in the Sgr B1 region. The results agree within the uncertainties with the obtained for twice solar metallicity using Parsec models. The only difference is a somewhat higher contribution of the youngest stellar population for lower metallicities (~7% and ~8% for 1.5 solar and solar metallicity, respectively).

**(2) KLFs bin width**. We analyzed the possible influence of the bin width on the derived SFH by repeating the process for new KLFs for the Sgr B1 and the control field, assuming half and double the previously computed bin width (0.03, and 0.12 mag for the Sgr B1 region, and 0.06, and 0.22 mag for the control field). The results are consistent with those obtained using Parsec models and a bin width of 0.06 mag.

To further assess whether the data binning influences our results, we resorted to cumulative luminosity functions, that allows us to eliminate the binning as a possible source of systematics. We built a cumulative Ks luminosity function for the Sgr B1 region and fitted it with Parsec theoretical models applying our MC simulation approach. Given the saturation of the data for de-reddened stars below $K_s = 8.5$ mag, we assumed the distance-modulus solution obtained for the standard KLF method previously explained, and computed the theoretical cumulative models without considering stars brighter than $K_s = 8.5$ mag that are not present in the real data and would significantly bias the results. This is a main disadvantage of this method, and implies to assume a-priory parameters for a proper fit. Moreover, cutting the bright end of the models requires to restrict the fit to a minimum value of $K_s = 8.75$ mag, to avoid a starting point of the models at the same magnitude as the real data. We assumed the ages and metallicity previously specified for Parsec models. We applied this technique to the Sgr B1 field and obtained that the results are compatible with the ones obtained by using the standard KLF fitting method, as shown in Fig. 9.

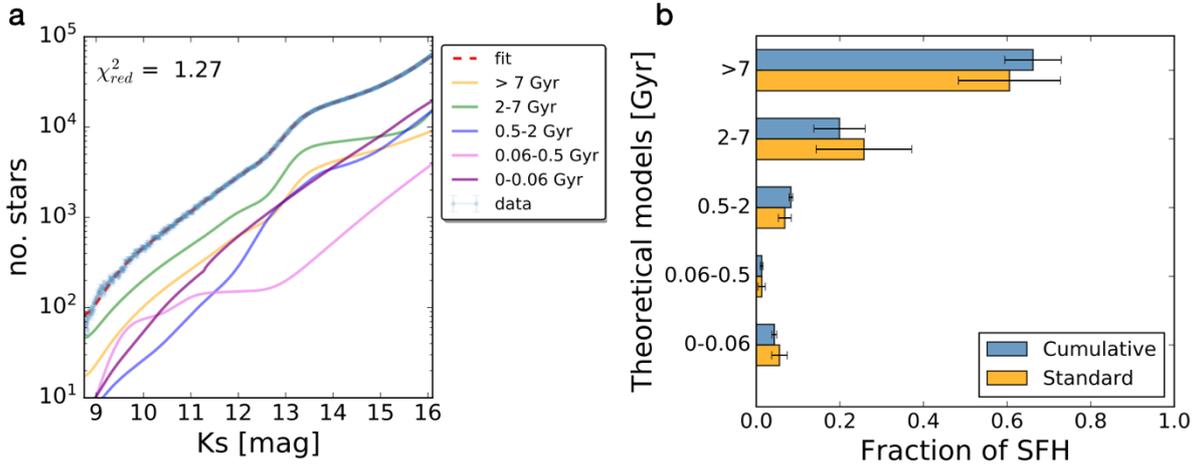

*Fig. 9: Analysis of the Sgr B1 cumulative KLF.* ***a***, *de-reddened and completeness corrected cumulative Ks luminosity function (in blue) obtained for one of the MC samples. The error bars show the 1σ uncertainties. The red dashed line depicts the best-fit model (Parsec models), whose reduced χ² is indicated in the Figure. Colored lines show the contribution of the different age bins.* ***b***, *SFH obtained by analyzing cumulative luminosity functions (in blue), and KLFs (in orange). The error bars indicate the 1σ uncertainties.*

**(3) Faint end of the KLF**. Given the different completeness of the Sgr B1 and the control fields (Sect. IV), we studied the influence of the deeper Sgr B1 photometry on the results. In this way, we repeated the analysis restricting the Sgr B1 KLF to the faint-end limit of the KLF of the control region (15.36 mag). We did not observe any significant difference within the uncertainties.

**(4) Bright end of the KLF**. We restricted the study of the KLFs to a Ks >8.5 mag limit to avoid any saturation problems related to the SIRIUS IRSF survey. Nevertheless, we also repeated the analysis of the Sgr B1 KLF considering a less conservative limit of Ks>7.5 mag. We concluded that there is not any significant variation on the results within the uncertainties.

**(5) Completeness solution**. We based our completeness corrections on artificial stars tests. To assess our completeness solution, we also estimated the completeness using an alternative approach based on the determination of the critical distance at which a star of a given magnitude can be detected around a brighter star[54]. This method is less accurate and constitutes a rough completeness estimation that mainly accounts for the completeness due to crowding. We obtained that the completeness is ~5-10 % lower than in the case of using the artificial stars test. We checked whether this different completeness solution influences our results, repeating the Sgr B1 KLF analysis by assuming the new completeness solution. We did not observe any variation within the uncertainties.

**(6) Different IMFs**. There is some evidence of a top-heavy IMF for the known young clusters at the Galactic center[32]. In this way, we repeated the analysis of the Sgr B1 KLF considering an IMF with α=1.8 using MIST models and twice the solar metallicity. We found that the SFH significantly changed for old ages, being the contribution from stellar models >7 Gyr shifted toward the 0.5-2 Gyr age bin. Nevertheless, the results for the youngest stellar bin did not change within the uncertainties, remaining the conclusion about the young stars unaffected. We believe that the age shift for the oldest stellar population is due to a probably necessary consideration of a stellar population enhanced in alpha elements[55], when assuming a top-heavy IMF, that was not considered due to the limitations of current models. Moreover, this alpha enhancement is required only for the young stars and might not be representative of the bulk of stars, making it probably necessary to use different metallicities and enhancement in alpha elements for stellar populations with different ages. On the other hand, we obtained that the mean chi2 computed for the described top-heavy models is significantly higher than for the standard case of using twice solar metallicity without enhancement in alpha elements. In any case, we conclude that the results for the youngest stellar population are robust.

**(7) Unresolved stellar multiple systems.** We expect a significant number of unresolved multiple stellar systems affecting the KLF, given the results obtained for local stellar populations[56]. To check the impact of stellar multiplicity in our results, we used the SPISEA python package[57] to compute theoretical luminosity functions accounting for unresolved multiple systems. We defined the multiplicity fraction, the companion star frequency, and the mass ratio between the multiple systems components using the standard SPISEA parameters, based on observations of young clusters[58] (<10 Myr), that correspond with the young stellar population that we find in our analysis. We computed theoretical KLFs using MIST models[24,25,26] (implemented in the SPISEA package), assuming around twice solar metallicity, and a slightly different age range due to the limitation of the models in SPISEA. In this way, we used 13 theoretical models with the following ages: 10, 8, 6, 3, 1.5, 1, 0.4, 0.2, 0.1, 0.04, 0.02, 0.01, and 0.005 Gyr. We applied our method to fit the KLFs corresponding to the Sgr B1 region and the control field. We restricted the bright end of the KLFs to $K_s = 9$ mag, due to the limitations of the models. The results are consistent with our findings for both, the Sgr B1 and the control regions.

**VI Theoretical models**

We used Parsec models[21,22,23] (version 1.2S+COLIBRI S37) as a main tool to derive the SFH of the analyzed regions. These models are designed to build synthetic stellar populations, considering a solar-scaled metal mixture. Parsec includes from pre-main-sequence stars to the thermally pulsing asymptotic giant branch, and considers a wide range of metallicities and ages. We sampled the age range to cover the main features visible in the KLF in the analyzed magnitude range. In particular, the KLF changes slowly for old stellar populations (> 7 Gyr), but more rapidly for younger ages (< 2 Gyr), where we increased the number of stellar models considered. In this way, we are able to reconstruct the SFH of a given stellar population by assuming a linear combination of the chosen theoretical models. Figure 10 shows the 14 theoretical models used.

To assess the results and analyze possible sources of systematic uncertainties due to the construction of the models, we repeated our analysis using MIST models[24,25,26]. They constitute a fully independent set of self-consistent models with solar-scaled abundance ratios, and extend across all evolutionary phases for all relevant stellar masses. We used a slightly different age range to properly cover the variation of the main features of the KLF, sampling the whole age-space parameter. This allows us to cover the change of brightness for RC stars around ~1 Gyr[ref4,44], that appears for slightly different ages between Parsec and MIST models (1.5 and 1 Gyr, respectively).

Given the smooth transition between models with similar ages, some degeneracy is expected. To decrease this potential degeneracy, we defined 5 larger age bins that include stellar populations with similar ages, and consider the ages defined for both, Parsec and MIST models.

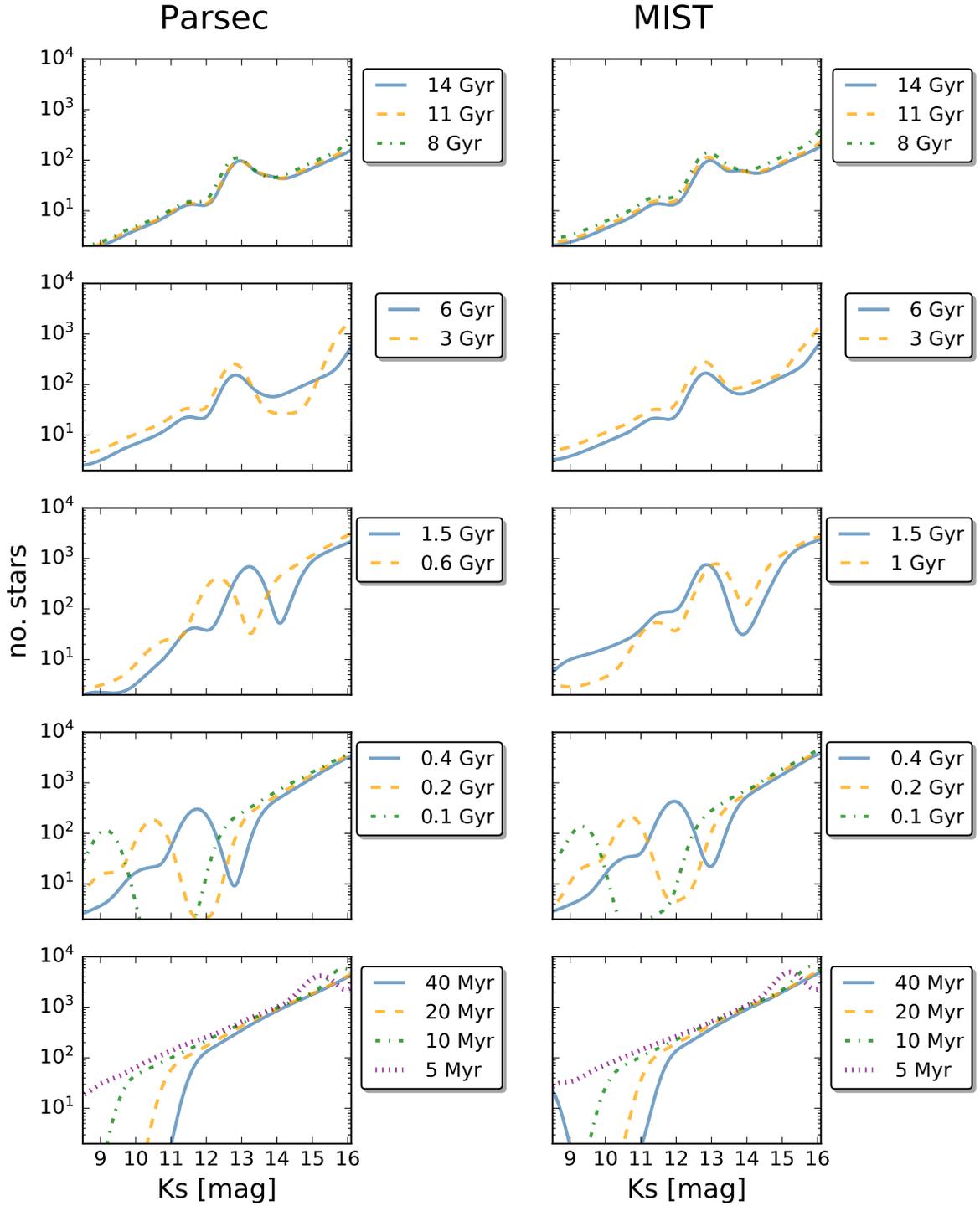

*Fig. 10: Synthetic stellar populations. Theoretical Parsec (left column), and MIST (right column) models used for the derivation of the SFHs. The models in each row are combined to create each of the 5 final age bins to decrease potential degeneracies between models with similar ages.*

## VII Youngest stellar models

Our analysis was specifically designed to properly detect the presence of young stellar populations. Therefore, we significantly increased the frequency of theoretical models contributing to the linear combination for ages <20 Myr (three models were considered: 5, 10, and 20 Myr). Given that there is not any stellar over-density that may be indicative of the presence of non-dissolved clusters in the Sgr B1 region, we chose the youngest model (5 Myr) in agreement with stars with ages slightly older than the known young clusters Arches and Quintuplet[32,35].

Figure 11 shows a comparison between stellar models in the range of 1-20 Myr to assess that the chosen models are a good reference for all the possible young ages. We conclude that there is a smooth transition between the different stellar populations and the 5, 10, and 20 Myr models represent a good choice to properly cover this range of ages.

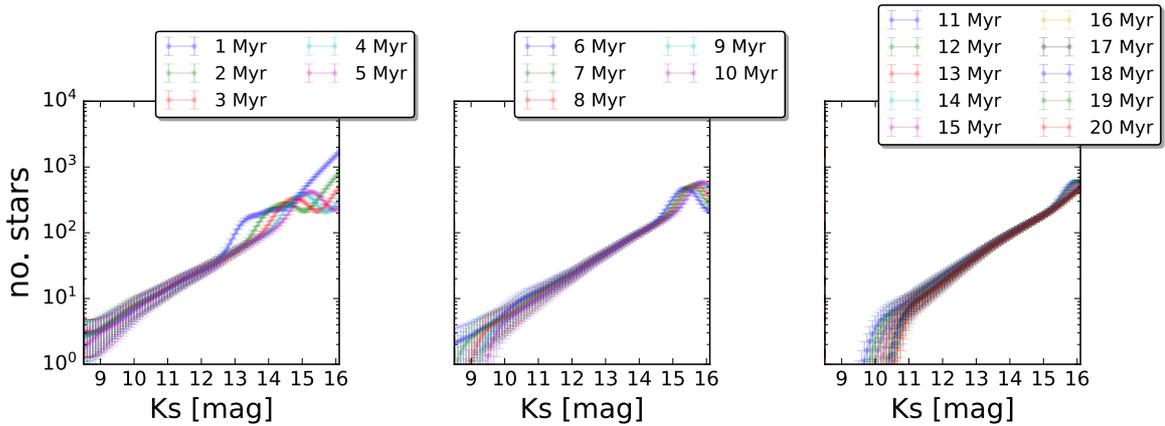

*Fig. 11: Theoretical Parsec models with twice solar metallicity and ages between 1 and 20 Myr in steps of 1 Myr. The ages shown in each panel are indicated in the legend. The stellar populations are plotted in a way that each panel contains one of the reference models assumed for the model fitting (5, 10, and 20 Myr, for panels **a**, **b**, and **c**, respectively). Each model contains a total mass of $10^5$ $M_\odot$, in agreement to the results for the youngest stellar population detected. The error bars show $1\sigma$ uncertainties that are estimated in agreement with those of real data. The Ks range corresponds to the KLF fitting range.*

## VIII Total mass estimation

We estimated the total stellar mass in the Sgr B1 region using the results obtained by fitting Parsec models. For this, we scaled the bin width of the KLF to the theoretical models' one and computed the stellar mass for each of the MC samples by combining the contributions of all the models in a given fit. The final stellar mass was obtained averaging over the results for each of the MC samples, where the associated uncertainty is the standard deviation of the mass distribution. The obtained value refers to the mass of the stellar population initially born.

## IX Region of intense hot dust emission

We carried out a dedicated analysis of the central region of Sgr B1, where the presence of hot dust emission is more intense than in the surrounding area based on a Spitzer 4.5 *μm* image (Fig. 1). We created a KLF and fitted it following the previously explained procedure. Figure 12 shows the obtained results. We found that the contribution from the youngest stellar population is higher than when considering the whole Sgr B1 region analyzed, suggesting an excess of young stars in this region. Moreover, we analyzed the contribution of the 5-10 Myr and the 10-20 Myr models to each of the MC samples (Fig. 3) from both, the whole Sgr B1 region and the region of intense hot dust emission. We found that the contribution from the 5-10 Myr stellar population is significantly higher in the region of intense hot dust emission (6±1 % versus 2±2 %). We also found that the contribution from stars in the age bin between 0.5-2 Gyr is more important, caused by a more significant contribution from the MIST models (that is significantly higher than the Parsec models' contribution) to the final result.

To assess the results, we analyzed the variation of the contribution of the youngest stellar bin when considering potential sources of systematic uncertainties, as we did in Sect. V for the whole Sgr B1 region. We concluded that the contribution of the youngest stellar population, and thus the detection of one or several dissolved clusters, is unaffected when considering: different bin widths of the KLF, variations of the faint and bright ends of the KLF, a top-heavy IMF, and a different completeness solution. On the other hand, we measured a somewhat higher contribution of the youngest stellar bin when considering solar metallicity (~14% of the total stellar mass is due to the youngest stellar population), and 1.5 solar metallicity (~10% of the stellar mass due to the youngest stellar population). Therefore, we concluded that the detection of a significant contribution (≳ 7 % of the stellar mass) of the youngest stellar population is robust.

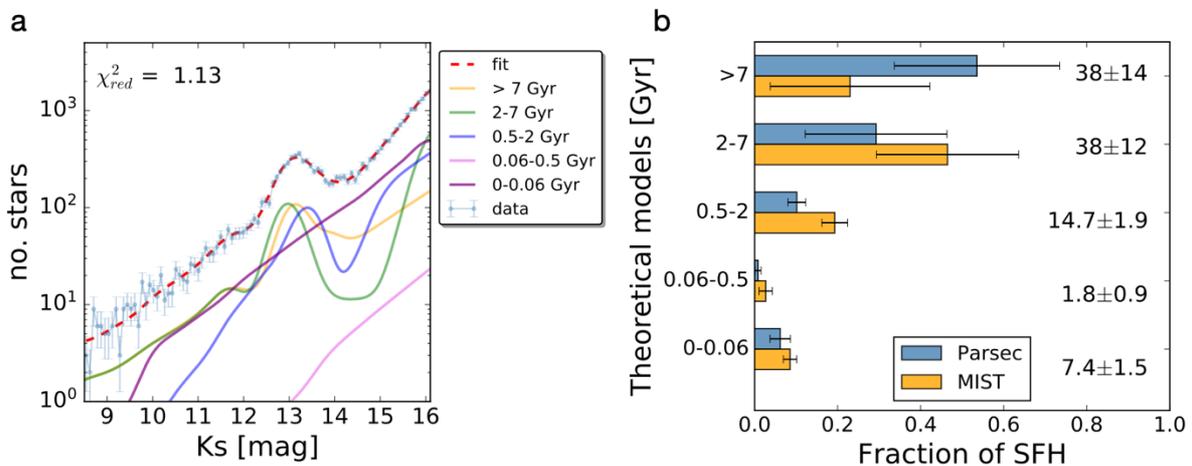

*Fig. 12: Analysis of the region of intense hot dust emission in the Spitzer 4.5 μm image (Fig. 1). **a**, de-reddened and completeness corrected KLF (in blue). The error bars show the 1σ uncertainties. The red dashed line depicts the best-fit model (Parsec models), whose reduced $\chi^2$ is indicated in the Figure. Colored lines show the contribution of the different age bins. **b**, SFH results. The error bars indicate the 1σ uncertainties. The numbers in each panel indicate the percentage of total stellar mass per age interval computed as an average between Parsec and MIST models (see Methods V).*

To assess the age estimate of the youngest stellar population, we repeated our analysis using Parsec models with ages: 14, 11, 8, 6, 3, 1.5, 0.6, 0.4, 0.2, 0.04, 0.02, 0.01, 0.005, 0.002 Gyr, where we included a stellar population with an age of 2 Myr to test the influence of a younger stellar population in the modeling. The results are fully compatible within the uncertainties with the ones assuming the previous stellar ages. We found that the contribution from the 5-10 Myr stellar population is compatible with our results, whereas there is no contribution from the 2 Myr model. In spite of the necessity of spectroscopic follow-up for an accurate age estimation, this indicates that the age that we estimate for the youngest stellar population (5-10 Myr) is consistent, and reinforces a scenario where the youngest stellar population was not formed in-situ.

# X Tests with artificial SFHs

We assessed the reliability of the KLF fitting method using synthetic SFHs created with Parsec models assuming twice solar metallicity. We built 4 different SFHs considering scenarios with and without the presence of young stars (Fig. 13). We assumed a stellar mass of ~2.2 · $10^6$ $M_\odot$, similar to the one obtained when analyzing the region of intense hot dust emission (the region with lowest stellar mass and thus, the most challenging case). We simulated the uncertainties for each stellar bin in agreement with real uncertainties computing them as the square root of the number of stars in a given magnitude bin. Figure 13a shows the simulated KLFs and also how they present different relative contribution of their characteristic features that allows us to reconstruct the SFH via model fitting. We applied the same analysis as for real data, and used the same magnitude limits. We obtained that the method is able to recover all the simulated SFHs within the $1\sigma$ uncertainties (Fig. 13b).

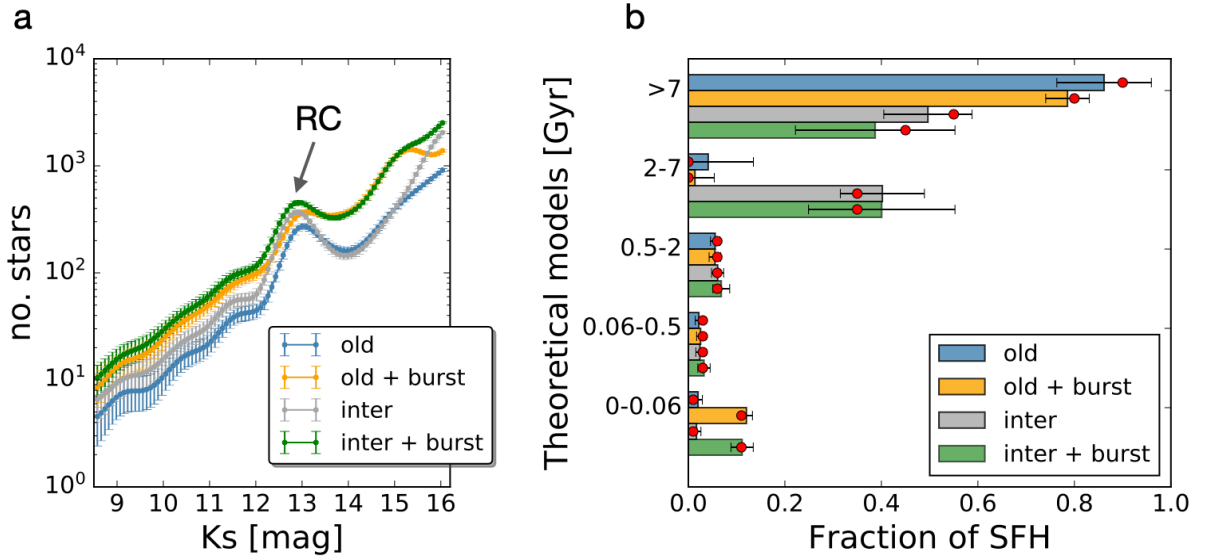

*Fig. 13: Analysis of synthetic stellar populations. **a**, Simulated KLFs with similar initial stellar mass corresponding to different SFHs (red dots in panel b). The error bars show the $1\sigma$ uncertainties, that were estimated assuming Poisson errors. The position of the red clump (RC) feature is indicated by an arrow. **b**, recovered SFHs after applying the model fitting technique. The error bars depict $1\sigma$ uncertainties.*

# XI Comparison with the Quintuplet cluster

We compared the results obtained for the region of intense hot dust emission with a same size region containing the Quintuplet cluster (F10 field of the GALACTICNUCLEUS survey[13], see Sect. I). Figure 14 shows the results obtained after applying our KLF fitting technique using Parsec and MIST models. We observed that the uncertainties are higher than for the case of the Sgr B1 region with intense hot dust emission. This is due to the lower data completeness in this region that limits the faint end of KLF to Ks~15 mag (~1 mag lower than for the Sgr B1 region). We estimated that the contribution of the very young stars considering the 5 + 10 Myr stellar models (the estimated age of the Quintuplet cluster is ~5 Myr) is 1.3±0.9 % of the total stellar mass in the region. This means a young stellar mass of (6.8±0.5) · $10^4$ $M_\odot$, that is of the same order of magnitude that the estimated mass for the Quintuplet cluster (~$10^4$ $M_\odot$), revealing its presence in the analyzed region and indicating that the method is capable of identifying young stars in a given field. A more precise measurement of the total young stellar mass would require deeper photometry to better constrain the faint end of the KLF.

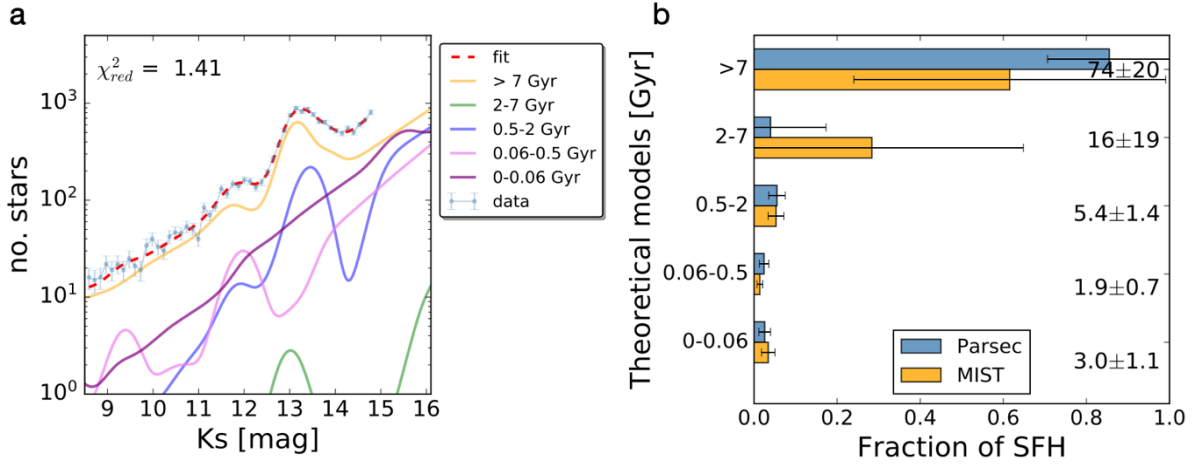

*Fig. 14: Analysis of the Quintuplet cluster region. **a**, de-reddened and completeness corrected KLF (in blue). The error bars show the 1σ uncertainty. The red dashed line depicts the best-fit model (Parsec models), whose reduced $\chi^2$ is indicated in the Figure. Colored lines show the contribution of the different age bins. **b**, SFH results. The error bars indicate the 1σ uncertainties. The numbers in each panel indicate the percentage of total stellar mass per age interval computed as an average between Parsec and MIST models (see Methods V).*

**Acknowledgements**

This work is based on observations made with ESO Telescopes at the La Silla Paranal Observatory under program ID 091.B-0418. We thank the staff of ESO for their great efforts and helpfulness. F. N.-L. acknowledges the sponsorship provided by the Federal Ministry for Education and Research of Germany through the Alexander von Humboldt Foundation. R.S. acknowledges financial support from the State Agency for Research of the Spanish MCIU through the "Center of Excellence Severo Ochoa" award for the Instituto de Astrofísica de Andalucía (SEV-2017-0709). R.S. acknowledges financial support from national project PGC2018-095049-B-C21 (MCIU/AEI/FEDER, UE), and financial support from grant P20-00753 awarded by Junta de Andalucía (Autonomical Government of Andalusia, Spain). N.N. gratefully acknowledges support by the Deutsche Forschungsgemeinschaft (DFG, German Research Foundation) Project-ID 138713538 SFB 881 ("The Milky Way System", subproject B8). F. N.-L. thanks Diederik Kruijssen, Adam Ginsburg, and Ashley Barnes for useful discussion.

**Author contributions**

F. N.-L. led and carried out the analysis, reduced the data, and wrote the manuscript. R. S. is the PI of the GALACTICNUCLEUS project and produced the MIST KLFs, and N. N. and R.S. contributed to the interpretation and the discussion of the results, and helped organizing and editing the manuscript in its final version.

**Competing Interests**

The authors declare no competing interests.